\documentclass[twocolumn,english,prd,superscriptaddress,nofootinbib,preprintnumbers]{revtex4}

\makeatletter

\providecommand{\tabularnewline}{\\}

\usepackage{amssymb}
\usepackage{graphicx}
\usepackage{dcolumn}
\usepackage{babel}
\makeatother

\begin{document}

\title{Challenges for scaling cosmologies}
%\title{An unifying approach for scaling cosmologies}
%\title{Unified approach for scaling cosmologies}

\author{Luca Amendola}

\email{amendola@mporzio.astro.it}

\affiliation{INAF/Osservatorio Astronomico di Roma, Via Frascati 33, I-00040 Monte
Porzio Catone, RM, Italy}

\author{Miguel Quartin}

\email{mquartin@if.ufrj.br}

\affiliation{Universidade Federal do Rio de Janeiro, Instituto de F{\'i}sica, CEP
21941-972, Rio de Janeiro, RJ, Brazil}

\author{Shinji Tsujikawa}

\email{shinji@nat.gunma-ct.ac.jp}

\affiliation{Department of Physics, Gunma National College of Technology, Gunma
371-8530, Japan}

\author{Ioav Waga}

\email{ioav@if.ufrj.br}

\affiliation{Universidade Federal do Rio de Janeiro, Instituto de F{\'i}sica, CEP
21941-972, Rio de Janeiro, RJ, Brazil}

\date{\today{}}

\begin{abstract}
A cosmological model that aims at solving the coincidence problem should show that
dark energy and dark matter follow the same scaling solution from some time onward.
At the same time, the model should contain a sufficiently long matter-dominated
epoch that takes place before acceleration in order to guarantee a decelerated epoch
and structure formation. So a successful cosmological model requires the occurrence
of a sequence of epochs, namely a radiation era, a matter-dominated era and a final
accelerated scaling attractor with $\Omega_{\varphi}\simeq0.7$. In this paper we
derive the generic form of a scalar-field Lagrangian that possesses scaling
solutions in the case where the coupling $Q$ between dark energy and dark matter is
a free function of the field $\varphi$. We then show, rather surprisingly, that the
aforementioned sequence of epochs cannot occur for a vast class of generalized
coupled scalar field Lagrangians that includes, to our knowledge, all scaling models
in the current literature.
\end{abstract}
\maketitle

\def\vp{\varphi}

\section{Introduction}

The unexpected discovery of the accelerated expansion of the universe opened a
Pandora's box of new issues and questions. Many of these are related to the nature
of dark energy and to its role within the particle physics model (see
Ref.~\cite{Reviews} for reviews). Other questions arise because of the so-called
coincidence problem: why two components that are completely unrelated and scale with
time in a different way, namely dark energy and matter, appear to have roughly the
same energy density just now and only now (``now'' here means within the last one or
two e-folding times).

It is possible that once we know the fundamental nature of dark energy the problem
of coincidence will be automatically and naturally explained. On the other hand, the
reverse could be true as well: understanding the origin of the coincidence could
shed light on the nature of dark energy and its relation to the rest of the world.
This is the footpath that we intend to pursue in this paper.

This work rests on a fundamental assumption: a complete solution of the coincidence problem
requires that dark energy and matter follow the same evolution with time, at least from some
time onward. For otherwise it is clear that the occurrence of coincidence will always depend
on the initial conditions of the system: changing the ratio dark energy/matter at some initial
time will always imply a displacement in time of the coincidence epoch. In other words, we can
explain the coincidence only if we show that it is not a coincidence at all, but rather that
energy and matter always (or from some time onward) shared a similar fraction of the total
budget. From a phase-space perspective, explaining the coincidence requires showing that our
\emph{present} universe has already attained its final attractor solution. A solution in which
matter and dark energy densities are both finite and have a constant ratio is denoted in
literature a \emph{scaling} solution \cite{Fer,CLW,scaling} or, if stable and accelerated, a
\emph{stationary} solution~\cite{dtv}.

Once one accepts this assumption, then it follows the first immediate consequence:
if we require that the energy density of dark energy is proportional to that of
matter (i.e., $\rho_{\rm DE}/\rho_{m}={\textrm{const}}$) and at the same time we
require that the dark energy equation of state parameter is less than $-1/3$ to get
an accelerated expansion, then one needs either to assume that matter has the same
negative equation of state as dark energy or that there is an interaction between
the two components, so that $\rho_{m}$ does not scale as $a^{-3}$. The first
possibility is clearly to be ruled out because such a modified matter equation of
state would profoundly affect the growth of perturbations. In fact, for any good
model it is not enough to require present acceleration: we also need the universe to
pass through a decelerated matter dominated epoch in the past in order to have a
well-behaved epoch of structure formation. A successful cosmological model should
therefore admit for a sequence of epochs: a radiation era, a sufficiently long
matter dominated era and a final stable accelerated scaling solution. This paper
aims at searching for such a ``good scaling cosmology''.

The assumption of a stable accelerated scaling solution requires therefore the
existence of an interaction between dark energy and matter \cite{coupled}. If dark
energy is modeled as a scalar field then the interaction with matter has to be a
scalar force additional to gravity. That is, our model has to be a scalar-tensor
gravitational theory \cite{scalartensor} or, equivalently, an Einsteinian theory
with an explicit coupling between matter and field. These models have been studied
for many times in the past and several important properties have been discussed
\cite{couplepapers,BNST,Tsujikawa06}. It is known, for instance, that a standard
scalar-tensor model with an exponential potential has a stable scaling solution and
that the scaling solution can be accelerated \cite{wet}. However, in this case, it
is possible to show that no matter phase precedes the acceleration. In other words,
after the radiation dominated era the system enters directly the accelerated regime.
This is in contrast with observations, as it has been shown in Ref.~\cite{coupled}.

The fact that the simplest case does not work is the main motivation for us to look
further. A simple generalization of the scalar field Lagrangian is to consider the
so-called $k$-essence Lagrangian \cite{kes}, such that the Lagrangian is a function
$p(X,\varphi)$ of the field $\vp$ and of the kinetic term
$X=-(1/2)g^{\mu\nu}\partial_{\mu}\vp\:\partial_{\nu}\vp$. This form is relatively
simple, being still second-order in the field, and has been already investigated for
several times. We note that this type of Lagrangian also covers a wide variety of
dark energy models such as quintessence \cite{quin}, tachyons \cite{tachyon},
phantoms \cite{phantom} and (dilatonic) ghost condensates \cite{dilaton,PT}. If
$\partial p/\partial X<0$, it has been shown that a phantom behavior ($w_{\rm
DE}<-1$) occurs \cite{BNST,Amendola06} and that matter feels a repulsive scalar
force \cite{Amendola04}. Moreover, it has been shown also that if the system
contains a scaling solution then $p$ can be cast in the form \cite{PT,TS}
\begin{equation}
    p(X,\vp)=Xg(Y)\,,\label{pt}
\end{equation}
 where $g(Y)$ is any function of the argument $Y=Xe^{\lambda\varphi}$
with $\lambda$ being a constant. For instance, $g(Y)=1-c/Y$ is in fact the standard
Lagrangian with an exponential potential ($p=X-ce^{-\lambda\varphi}$). The above
form for $p$ was shown to be valid for uncoupled dark energy and for the case in
which the coupling
\begin{equation}
    Q\equiv\frac{-1}{\rho_{m}\sqrt{-g_{M}}}\frac{\delta
    S_{m}}{\delta\vp}\,,\label{Qdef}
\end{equation}
 is a constant (here $S_{m}$ is the action for matter and $g_{M}$
is the metric determinant).

In this paper we perform a search of a good scaling cosmology in three steps. First,
we show that the Lagrangian (\ref{pt}) extends also to the case of variable coupling
$Q(\varphi)$ up to a field redefinition. This is an interesting
result in itself
since it unifies some sporadic results obtained in different ways in literature
(e.g. \cite{Chinese}). Then, assuming as a model for $g$ a polynomial
\begin{equation}
    g=\sum_{n}c_{n}Y^{-n}\,,
\end{equation}
with both positive and negative integer powers of $n$, we derive the critical points
of the system. Finally, we show that within this class of models there is no way to
obtain a sequence of a matter phase followed by a stable accelerated scaling
solution. When a kinetic scaling matter-dominated era exists, this stage is
generally followed by a scalar-field dominated attractor ($\Omega_{\vp}=1$) instead
of an accelerated scaling attractor. Although our proof of absence of two scaling
regimes does not extend to any possible $g(Y)$, we believe that it seriously
undermines the real possibility of realizing such an ideal cosmology. This negative
result opens the challenge: is there \emph{any} case in which a successful sequence
can be realized? In the final section we will comment on the possibility of
obtaining a good scaling cosmology with a fractional power Lagrangian $g=c_{0}-c\,
Y^{-u}$, in which $0<u<1$. Nevertheless, we leave a more complete study of this,
perhaps very exotic, case to future work.

Beside being a way to approach the problem of coincidence, a scaling cosmology also
provides us with a useful alternative to standard dark energy scenarios. The
behavior of the background cosmology and of its linear perturbations is in fact
radically different in scaling cosmology with respect to most other models. Let us
just mention three basic differences (see e.g., Refs.~\cite{dtv,agp,APT}). First, in
a scaling cosmology the acceleration could start at any epoch in the past. Second,
the perturbations can keep growing even during the accelerated regime. Third, the
amount of dark energy does not necessarily become negligible at high redshifts. All
three features radically distinguish scaling cosmologies from usual dark energy
models which only focus on dark energy itself and not on its relation with matter.
As such, scaling cosmologies may serve as a useful testing ground for observations.

Before passing to the actual calculations, we should spend a note on the local
gravity constraints on scalar forces. In principle, the coupling we introduce is
severely constrained by local gravity experiments on scalar-tensor theories.
However, these can be escaped at least in three ways. First, by designing a
potential with a large mass and, consequently, a short interaction range
\cite{cham}. Second, by building a model that happens to satisfy the constraints
now, but not in the past. Third, by assuming that the baryons are actually uncoupled
to the scalar field~\cite{coupled}. The first two solutions change the potential and
affect the global evolution and therefore will not in general satisfy the requisite
for a cosmology that solves the coincidence problem. The third case on the contrary
can be implemented without affecting the potential of the scalar field.

In general, a component of uncoupled baryons can dominate in the past, even if their
abundance now is very small \cite{dtv}. In this case a matter phase does exist, but
it is a \emph{baryonic} matter epoch instead of a dark matter epoch. This raises
many problems on its own. For instance, the baryonic perturbations are almost erased
on small scales due to the coupling to radiation and therefore, without the support
from dark matter, would hardly grow to the observed amplitude; moreover, the
baryonic era would finish early in the past, at redshifts quite larger than 1 and
the subsequent accelerated regime would be too long to be in fair agreement with
both the supernovae experiments (although here the discrepancy is marginal, see
Ref.~\cite{agp}) and with the integrated Sachs-Wolfe effect \cite{dtv}. In any case,
if a standard (dark) matter phase exists, the baryons would never dominate, as we
will argue later on. Hence the search for a ``good scaling cosmology'' can simply
neglect the small baryon component and this is what will be done in the present
work. Finally, since the two matter components have a different coupling, one has to
choose a physical frame in which baryons are conserved (otherwise particle masses
will be time-varying) but dark matter is not. We will work therefore in this frame,
which is the so-called Einstein frame.

\section{The General Lagrangian for Scaling Solutions with Arbitrary Coupling}

We shall derive here the general Lagrangian admitting scaling solutions in the case
where the coupling between dark energy and dark matter depends upon the field $\vp$.
This is the generalization of the works \cite{PT,TS}. Let us start by considering
the following action, written in the Einstein frame:
\begin{equation}
    S=\int{\textrm{d}}^{4}x\sqrt{-g_{M}}\left[\frac{M_{P}^{2}}{2}R+p(X,\varphi)\right]+
    S_{m}[\varphi,\psi_{i},g_{\mu\nu}],\label{eqn:action}
\end{equation}
 where $M_{P}$ is the reduced Planck mass, $R$ is the Ricci scalar,
$\vp$ is a scalar field, $X=-(1/2)g^{\mu\nu}\partial_{\mu}\vp\:\partial_{\nu}\vp$
and $\psi_{i}$ are the various matter fields. Notice that we allow for an arbitrary
coupling between the matter fields and the scalar field $\vp$. As mentioned in the
Introduction, in order to cope with current observations, we assume that $\vp$
couple only to dark matter. We will also suppose that the dark matter component
dominates over any other baryonic form of matter.

We are interested in cosmological scaling solutions in a spatially flat
Friedmann-Robertson-Walker (FRW) background metric with a scale factor $a(t)$:
\begin{equation}
    {\textrm{d}}s^{2}=-{\textrm{d}}t^{2}+a^{2}(t){\textrm{d}}\mathbf{x}^{2}.
    \label{eqn:FRW}
\end{equation}
 Friedmann equation in Einstein gravity is given by
\begin{equation}
    3H^{2}=M_{P}^{-2}\rho_{\rm T}\,,\label{Hubble}
\end{equation}
where $M_{P}^{-2}=8\pi G$ with $G$ being the gravitational constant, and $\rho_{\rm
T}$ is the total energy density of the universe. In what follows we shall set
$M_{P}=1$.

Our focus will be on solutions with constant equation of state parameter
$w_{\vp}\equiv p(X,\vp)/\rho_{\vp}$ in the scaling regime and in which the universe
is filled only by two components: a barotropic fluid (such that $w_{m}\equiv
p_{m}/\rho_{m}$) and the scalar field $\vp$. Rewriting the Klein-Gordon equation for
the field $\vp$ (in the above metric) in terms of its energy density,
$\rho_{\vp}=2X\partial p/\partial X-p$, one gets \cite{PT}
\begin{equation}
    \frac{{\textrm{d}}\rho_{\vp}}{{\textrm{d}}N}+3(1+w_{\vp})\rho_{\vp}=-Q\rho_{m}
    \frac{{\textrm{d}}\vp}{{\textrm{d}}N}\,,\label{eqn:conserv-eqn-field}
\end{equation}
 where $N\equiv\ln a$ and $Q$ is defined by Eq.~(\ref{Qdef}).

If one starts from scalar-tensor theories \cite{scalartensor} or a mass-varying
neutrino scenario \cite{neutrino}, we have
\begin{equation}
    \frac{{\textrm{d}}\rho_{\vp}}{{\textrm{d}}N}+3(1+w_{\vp})\rho_{\vp}=-\tilde{Q}(1-3w_{m})
    \rho_{m}\frac{{\textrm{d}}\vp}{{\textrm{d}}N}\,.
\end{equation}
 This case reduces to Eq.~(\ref{eqn:conserv-eqn-field}) if one relates
the coupling $\tilde{Q}$ to the coupling $Q$ as $\tilde{Q}(1-3w_{m})=Q$. In what
follows we shall derive the condition for the existence of scaling solutions by
using Eq.~(\ref{eqn:conserv-eqn-field}). Note that the energy density of a
barotropic fluid satisfies
\begin{equation}
    \frac{{\textrm{d}}\rho_{m}}{{\textrm{d}}N}+3(1+w_{m})\rho_{m}=Q\rho_{m}
    \frac{{\textrm{d}}\vp}{{{\textrm{d}}N}}.\label{eqn:conserv-eqn-m}
\end{equation}

We shall define the fractional densities of $\rho_{\vp}$ and $\rho_{m}$ as
\begin{equation}
    \Omega_{\vp}\equiv\frac{\rho_{\vp}}{3H^{2}}\,,\quad\Omega_{m}\equiv
    \frac{\rho_{m}}{3H^{2}}\,.
\end{equation}
 These satisfy $\Omega_{\vp}+\Omega_{m}=1$ from Eq.~(\ref{Hubble}).
Scaling solutions are characterized by the condition
$\rho_{\vp}/\rho_{m}={\textrm{const}}$, in which case $\Omega_{\vp}$ is a constant.
Using these relations, together with Eqs.~(\ref{eqn:conserv-eqn-field}) and
(\ref{eqn:conserv-eqn-m}), and following the procedure of Refs.~\cite{PT,TS} but
dropping the assumption of a constant coupling $Q(\vp)$, one finds the relations
\begin{equation}
    \frac{{\textrm{d}}\ln
    p_{\varphi}}{{\textrm{d}}N}=\frac{{\textrm{d}}\ln\rho_{\vp}}{{\textrm{d}}N}=
    \frac{{\textrm{d}}\ln\rho_{m}}{{\textrm{d}}N}=-3(1+w_{\rm eff})\,,
    \label{eqn:drho-dN}
\end{equation}
 and
\begin{equation}
    \frac{{\textrm{d}}\vp}{{\textrm{d}}N}=\frac{3\Omega_{\vp}}{Q(\vp)}(w_{m}-w_{\vp})
    \propto\frac{1}{Q(\vp)}\,,\label{eqn:dphi-dN}
\end{equation}
 where we introduced an effective quantity:
\begin{equation}
    w_{\rm eff}\equiv
    w_{m}\Omega_{m}+w_{\vp}\Omega_{\vp}.\label{eqn:def-w-eff}
\end{equation}

From these equations and the definition of $X$, one arrives at
\begin{equation}
    2\,X=H^{2}\left(\frac{{\textrm{d}}\vp}{{{\textrm{d}}N}}\right)^{2}\;\;\propto\;\;
    \frac{\rho_{\vp}}{{Q^{2}}}\;\;\propto\;\;\frac{p(X,\vp)}{Q^{2}},
    \label{eqn:X-proportional-to}
\end{equation}
 and thus
\begin{equation}
    \frac{{\textrm{d}}\ln X}{{\textrm{d}}N}=-3(1+w_{\rm eff})-2\frac{{\textrm{d}}\ln
    Q}{{\textrm{d}}N}\,.\label{eqn:dlnX-dN}
\end{equation}
 Making use of Eqs.~(\ref{eqn:drho-dN}), (\ref{eqn:dphi-dN}) and
(\ref{eqn:dlnX-dN}) we arrive at the following generalized ``master equation'' for
the Lagrangian $p(X,\vp)$:
\begin{equation}
    \left[1+\frac{2}{\lambda\,
    Q^{2}}\frac{{\textrm{d}}Q(\vp)}{{{\textrm{d}}\vp}}\right]\frac{\partial\ln
    p}{\partial\ln X}-\frac{1}{\lambda Q}\frac{\partial\ln
    p}{\partial\vp}=1,\label{eqn:gen-master-eqn}
\end{equation}
where
\begin{equation}
    \lambda\,\equiv\,\frac{1+w_{m}-\Omega_{\vp}(w_{m}-w_{\vp})}{\Omega_{\vp}
    (w_{m}-w_{\vp})}.\label{eqn:def-lambda}
\end{equation}
Equation (\ref{eqn:gen-master-eqn}) reduces to the one found in Ref.~\cite{PT} when
$Q(\vp)$ is constant.

Solving Eq. (\ref{eqn:gen-master-eqn}) one gets:
\begin{equation}
    p(X,\vp)=XQ^{2}(\vp)\; g\left(XQ^{2}(\vp)\,
    e^{\lambda\psi(\vp)}\right),\label{eqn:solution-for-p}
\end{equation}
 where $g$ is an arbitrary function and
\begin{equation}
    \psi(\vp)\equiv\,\int^{\vp}Q(\xi){\textrm{d}}\xi.\label{eqn:def-of-psi}
\end{equation}
 See Appendix A for the derivation of Eq.~(\ref{eqn:solution-for-p}).
In the case of constant coupling both $Q^{2}$ terms in
Eq.~(\ref{eqn:solution-for-p}) can be absorbed in the definition of $g$, so our
solution reduces to that in Ref.~\cite{PT}. In a nutshell, what
Eq.~(\ref{eqn:solution-for-p}) means is that any Lagrangian that allows scaling
solutions with constant $w_{\vp}$ can always be cast in the above form by a
convenient field redefinition. The standard kinetic case corresponds therefore to
$p=XQ^{2}-e^{-\lambda\psi}$. Another example is provided by a coupling $Q=1/\varphi$
as in Ref.~\cite{Chinese}. By using Eq.~(\ref{eqn:def-of-psi}) we find that the
$e^{\lambda\psi}$ term in Eq.~(\ref{eqn:solution-for-p}) is given by
$e^{\lambda\psi}=\varphi^{\lambda}$. In this case the Lagrangian
(\ref{eqn:solution-for-p}) becomes $p=Xg(X\varphi^{\lambda-2})/\varphi^{2}$.
When $\lambda=2$ this simplifies to $p=\bar{g}(X)/\varphi^{2}$, where $\bar{g}(X)\equiv
Xg(X)$ is an arbitrary function of $X$. This form of $p$ corresponds to the choice
given in Ref.~\cite{Chinese}.

Now let us make the following field redefinition: $\vp\rightarrow\psi(\vp)$, with
$\psi(\vp)$ as defined in (\ref{eqn:def-of-psi}). This in turn implies $X\rightarrow
X_{\psi}=X\, Q^{2}(\vp)$, and Eq.~(\ref{eqn:solution-for-p}) becomes:
\begin{equation}
    p(X_{\psi},\psi)=X_{\psi}\; g\left(X_{\psi}\, e^{\lambda\psi}\right)\,,\label{geLag}
\end{equation}
 which is the same functional form found in the constant coupling
scenario \cite{PT}. At the same time the relation between $S_{m}$ and the coupling
becomes
\begin{equation}
    1=\frac{-1}{\rho_{m}\sqrt{-g_{M}}}\frac{\delta S_{m}}{\delta\psi}\,.
\end{equation}
 We have thus shown that the case of a constant coupling ($Q=1$)
is the most general one. In other words, if one is interested in scaling solutions,
one can always work with a Lagrangian in the above form, no matter what kind of
coupling one has in mind.

In order to follow the current notation in the literature (e.g.
\cite{PT,TS}) we will use, instead of $\psi$, the field $\bar{\vp}(\psi)$ defined by
$\bar{\vp}(\psi)\,\equiv\, \psi / \bar{Q}$, where $\bar{Q}$ is a constant. In this
case one can absorb the $\bar{Q}$ term that appears in the exponential in the
argument of $g$ into the definition of~$\lambda$. Hence, in what follows, we shall
always consider the Lagrangian density (dropping the bars on $Q$ and $\vp$)
\begin{equation}
    p=X\, g(Xe^{\lambda\vp})\,,\label{geLag2}
\end{equation}
 where now $\lambda$ is given by
\begin{equation}
    \lambda\,\equiv\,Q
    \frac{1+w_{m}-\Omega_{\vp}(w_{m}-w_{\vp})}{{\Omega_{\vp}(w_{m}-w_{\vp})}}\,.
\end{equation}
 It is important at this stage to realize that the system is invariant
under a simultaneous change of sign of $Q$ and $\lambda$. We can therefore without
loss of generality consider only the case $\lambda>0$.

In the Appendix B we generalize the above results for a more general cosmological
background, in which Eq.~(\ref{Hubble}) is replaced by $H^{2}\propto \rho_{\rm
T}^{q}$. In subsequent sections, though, we shall restrict
the analysis to Einstein gravity ($q=1$).

\section{Phase Space Equations}

So far we have derived the most general Lagrangian that possesses scaling solutions.
In addition to scaling solutions there exist other fixed points for the
system~(\ref{geLag2}) characterized by $\Omega_{\vp}=1$. In what follows
we shall derive the autonomous equations taking into account radiation to find the
general behavior of the solutions. As we mentioned in the Introduction we are
interested in searching for a good scaling cosmology, namely a sequence composed of
a radiation epoch, a matter-dominated era and an accelerated scaling attractor.

Many results will be shown to hold for any $g(Y)$. However we carry out our search
assuming as a reference model a polynomial expansion in positive and negative
\emph{integer} powers:
\begin{eqnarray}
    g=c_{0}+\sum_{n>0}c_{n}Y^{-n}+\sum_{n'<0}c_{n'}Y^{-n'}\,,\label{eq:poly}
\end{eqnarray}
 where $c_{0}$, $c_{n}$ and $c_{n'}$ are constants. Note that if
we have $c_{0}=1$ and all other $c_{n}$, except $c_{1}$, being zero, our case
reduces to that of an ordinary scalar field with an exponential potential
\cite{CLW}.
%Note that we can always multiply the Lagrangian $p$ by a constant
%leaving the model unchanged up to a redefinition of $Q$: the constant $c_{0}$ can
%therefore be implicitly assumed to be either 0 or $\pm1$.
% {I deleted this sentence, but Luca may recover it if you want. by S.T.}

For the Lagrangian density (\ref{geLag2}) in the presence of pressureless dust and
radiation we obtain the following equations
\begin{eqnarray}
    &  & 3H^{2}=X(g+2g_{1})+\rho_{m}+\rho_{\rm rad}\,,\label{ba1}\\
    &  & 2\dot{H}=-\left[2X(g+g_{1})+\rho_{m}+\frac{4}{3}\rho_{\rm rad}\right]\,,
    \label{ba2}\\
    &  & \ddot{\vp}+3AH(g+g_{1})\dot{\vp}+\lambda X[1-A(g+2g_{1})]\nonumber \\
    &  & +AQ\rho_{m}=0,\label{ba3}
\end{eqnarray}
 where $A\equiv(g+5g_{1}+2g_{2})^{-1}$ and
\begin{equation}
    g_{n}\equiv Y^{n}\,\partial^{n}g/\partial Y^{n}.
\end{equation}
The speed of sound, $c_{s}$, is related to the quantity $A$ by \cite{PT,Tsujikawa06}
\begin{equation}
    c_{s}^{2}=A\,(g+g_{1})\,.
\end{equation}
 When $A^{-1}=0$ the speed of sound diverges. Hence no physically
acceptable evolution can cross the border $A^{-1}=0$.

In order to study the dynamics of the above system it is convenient to introduce the
following dimensionless quantities:
\begin{eqnarray}
    x=\frac{\dot{\vp}}{\sqrt{6}H}\,,~~~y=\frac{e^{-\lambda\vp/2}}{\sqrt{3}H}\,,~~~
    z=\frac{\sqrt{\rho_{\rm rad}}}{\sqrt{3}H}\,.
\end{eqnarray}
 Then $Y$ is written as $Y=x^{2}/y^{2}$. Equation (\ref{ba1}) gives
the following constraint equation
\begin{eqnarray}
    \Omega_{m}\equiv\frac{\rho_{m}}{3H^{2}}=1-\Omega_{\vp}-z^{2}\,,\label{constraint}
\end{eqnarray}
 where
\begin{eqnarray}
    \Omega_{\vp}=x^{2}(g+2g_{1})\,.\label{Omedef}
\end{eqnarray}
It is important to note that, in principle,
$\Omega_{\vp}$ could as well be negative. From Eq.~(\ref{ba2}) we find
\begin{eqnarray}
\frac{1}{H}\frac{{\textrm{d}}H}{{\textrm{d}}N}=-\frac{1}{2}(3+3gx^{2}+z^{2})\,.\label{dotH}
\end{eqnarray}

By using Eqs.~(\ref{ba3}), (\ref{constraint}) and (\ref{dotH}), we obtain the
autonomous equations:
\begin{eqnarray}
    \frac{{\textrm{d}}x}{{\textrm{d}}N} & = & \frac{3}{2}x\left[1+gx^{2}-2A(g+g_{1})+
    \frac{1}{3}z^{2}\right]\nonumber \\
    &  & +\frac{\sqrt{6}}{2}{\Big[}A(Q+\lambda)(g+2g_{1})x^{2}-\lambda x^{2}\nonumber \\
    &  & \;\;\;\qquad+\, Q A(z^{2}-1){\Big]},\label{auto1}\\
    \frac{{\textrm{d}}y}{{\textrm{d}}N} & = & \frac{y}{2}\left(3-\sqrt{6}\lambda
    x+3gx^{2}+z^{2}\right),\label{auto2}\\
    \frac{{\textrm{d}}z}{{\textrm{d}}N} & = &
    \frac{z}{2}\left(-1+3gx^{2}+z^{2}\right).\label{auto3}
\end{eqnarray}
 It is useful to notice the relations
\begin{eqnarray}
    \rho_{\vp}=X(g+2g_{1}),\quad w_{\vp}=\frac{g}{g+2g_{1}}\,,\label{rhowvp}
\end{eqnarray}
 and also
\begin{eqnarray}
    w_{\vp}=-1+\frac{2x^{2}}{\Omega_{\vp}}(g+g_{1})\,,\quad\Omega_{\vp}w_{\vp}=gx^{2}\,.
    \label{wphire}
\end{eqnarray}
 This means that $w_{\vp}>-1$ for $g+g_{1}>0$ and $w_{\vp}<-1$
for $g+g_{1}<0$. From Eq.~(\ref{dotH}) the effective equation of state parameter of
the system is given by
\begin{eqnarray}
    w_{\rm eff}=-1-\frac{2}{3}\frac{\dot{H}}{H^{2}}=gx^{2}+\frac{1}{3}z^{2}\,.\label{weffex}
\end{eqnarray}
 Then, in the absence of radiation ($z=0$), one has \linebreak[4] $w_{\rm eff}=
\Omega_{\vp}w_{\vp}$ {[}see Eq.~(\ref{wphire}){]}.

\section{Critical points}

In this section we shall derive the fixed points for the above autonomous system in
the absence of radiation ($z=0$). The critical
points for $z\neq0$ are irrelevant to our study. They are listed in the Appendix
C, for the sake of completeness. The fixed points for $z=0$ are derived by setting
${\textrm{d}}x/{\textrm{d}}N=0$ and ${\textrm{d}}y/{\textrm{d}}N=0$ in
Eqs.~(\ref{auto1}) and (\ref{auto2}). From Eq.~(\ref{auto2}) we find two distinct
classes of solutions, either for $3-\sqrt{6}\lambda x+3gx^{2}=0$ or for $y=0$. The
former case gives both scalar-field dominated and scaling
solutions~\cite{Tsujikawa06}. In fact in this case we have
\begin{eqnarray}
    x=\frac{\sqrt{6}(1+w_{\vp}\Omega_{\vp})}{2\lambda}\,,\label{xexpress}
\end{eqnarray}
 and, inserting this into Eq.~(\ref{auto1}) we find two cases:

\begin{itemize}
\item Point A: a scalar-field dominated (SFD) solution with
\begin{eqnarray}
    \Omega_{\vp}=1\,.\label{fixedi}
\end{eqnarray}

\item Point B: a scaling solution with
\begin{eqnarray}
    \Omega_{\vp}=-\frac{Q}{w_{\vp}(Q+\lambda)}\,.\label{scaome}
\end{eqnarray}
\end{itemize}
Let us remind that by definition a scaling solution corresponds to a situation in
which $\Omega_{\vp}$ equals neither $1$~nor~$0$.

The properties of the points A and B will be discussed in subsections A and B,
respectively. In subsection C we shall discuss the second class of solutions, in
which \linebreak[4] $y=0$. Since these solutions exist in the limit of a vanishing
potential, we denote them as \emph{kinetic solutions}.

\subsection{Point A: Scalar-field dominated solutions}

When $\Omega_{\vp}=1$, we have the following relations \cite{Tsujikawa06}
\begin{eqnarray}
    g(Y_A)=\frac{\sqrt{6}\lambda x_A-3}{3x_A^{2}}\,,~~~g_{1}(Y_A)=
    \frac{6-\sqrt{6}\lambda x_A}{6x_A^{2}}\,. \label{gre1}
\end{eqnarray}
 Specifying the model $g(Y)$ one obtains $Y_A$ and the fixed point
$(x_{A},y_{A})$ by using Eq.~(\ref{gre1}) and the relation \linebreak[4]
$Y=x^{2}/y^{2}$. From Eq.~(\ref{wphire}) we find
\begin{eqnarray}
    w_{\rm eff}=w_{\vp}=-1+\frac{\sqrt{6}\lambda}{3}x_{A}\,.\label{eqsta}
\end{eqnarray}
A general Lagrangian could have in principle many different classes
of the point A.
The number of such critical points is known by solving Eq.~(\ref{gre1}).

The stability of fixed points can be analyzed by considering linear perturbations
around them. This was carried out in Ref.~\cite{Tsujikawa06} for a general $g(Y)$
for positive values of $Q$ and $\lambda$. The eigenvalues of the matrix for
perturbations are given by
\begin{eqnarray}
    \mu_{+}=-3+\sqrt{6}(Q+\lambda)x_{A}\,, \quad
    \mu_{-}=-3+\frac{\sqrt{6}}{2}\lambda x_{A}\,.
\end{eqnarray}
The fixed point is a stable node if $\mu_+<0$
and $\mu_-<0$.
Allowing negative values of $Q$ as well,
the fixed point A is stable when
\begin{eqnarray}\label{stacon}
    \left\{
    \begin{array}{ll}
        x_{A}<\frac{\sqrt{6}}{2(Q+\lambda)}\,,& \;\mbox{if} \;\; Q>-\lambda/2\,, \\
        x_{A}<\frac{\sqrt{6}}{\lambda}\,,& \;\mbox{if} \;\; -\lambda<Q\leq-\lambda/2\,, \\
       \frac{\sqrt{6}}{2(Q+\lambda)}< x_{A}<\frac{\sqrt{6}}{\lambda}\,,& \;\mbox{if}
       \;\; Q<-\lambda\,.
    \end{array}
    \right.
\end{eqnarray}

For negative $x_{A}$, which corresponds to a phantom equation of state
($w_{\vp}<-1$) from Eq.~(\ref{eqsta}), the first two conditions in
Eq.~(\ref{stacon}) are automatically satisfied. Hence when $Q>-\lambda$ the phantom
fixed point  is always classically stable. On the other hand the stability of
non-phantom fixed points ($x_{A}>0$) depends upon the values of $Q$ and $\lambda$.
Since $x_{A}$ is given by $x_{A}=\lambda/[\sqrt{6}(g+g_1)]$, the stability condition
(\ref{stacon}) for non-phantom fixed points is expressed as
\begin{eqnarray}\label{stacon2}
    \left\{
    \begin{array}{ll}
        g+g_1>\lambda (Q+\lambda)/3\,,& \;\mbox{if} \;\; Q>-\lambda/2\,, \\
        g+g_1>\lambda^2/6\,,& \;\mbox{if} \;\; Q\leq-\lambda/2\,.
    \end{array}
    \right.
\end{eqnarray}

\subsection{Point B: Scaling solutions}

The scaling solution satisfies the relation (\ref{scaome}). Then
Eq.~(\ref{xexpress}) gives
\begin{eqnarray}
    x_{B}=\frac{\sqrt{6}}{2(Q+\lambda)}.\label{scalingB}
\end{eqnarray}
 We also obtain the following relations valid for all $g$ \cite{Tsujikawa06}:
\begin{eqnarray}
    &  & g(Y_{B})=-\frac{2Q(Q+\lambda)}{3}\,,\label{scaling1}\\
    &  & w_{\rm eff}=-\frac{Q}{Q+\lambda}\,,\label{scaling2}\\
    &  & w_{\vp}=-\frac{Q(Q+\lambda)}{Q(Q+\lambda)+3(g+g_{1})}\,,\label{scaling3}\\
    &  & \Omega_{\vp}=\frac{Q(Q+\lambda)+3(g+g_{1})}{(Q+\lambda)^{2}}\,.\label{scaling4}
\end{eqnarray}
Again, once the function $g(Y)$ is specified, one obtains $Y_{B}$ and
$y_{B}=|x_{B}|/\sqrt{Y_{B}}$ as a function of $Q,\lambda$. The condition for an
accelerated expansion corresponds to $w_{\rm eff}<-1/3$ and this gives us
\begin{eqnarray}
    Q>\lambda/2\quad\,\mbox{or}\quad\, Q<-\lambda\,,\label{con1}
\end{eqnarray}
which are again independent of the form of $g(Y)$. Note that the latter case
corresponds to the effective phantom ($w_{\rm eff}<-1$) as we see from
Eq.~(\ref{scaling2}).

The eigenvalues of the matrix for perturbations around the fixed point B are given
by \cite{Tsujikawa06}
\begin{eqnarray}
    \mu_{\pm}=\xi_1 \left[1 \pm \sqrt{1-\xi_2} \right]\,,
\end{eqnarray}
where
\begin{eqnarray}
    & &\xi_1=-\frac{3(2Q+\lambda)}{4(Q+\lambda)}\,, \\
    & & \xi_2=\frac{8(1-\Omega_\vp)(Q+\lambda)^3
    [\Omega_{\vp} (Q+\lambda)+Q]}{3(2Q+\lambda)^2}A\,.
\end{eqnarray}
This point is stable if $\xi_{1}<0$ and $\xi_2>0$. We find that negative $\xi_{1}$
corresponds to $Q>-\lambda/2\,$ or $\,Q<-\lambda$. Hence when the condition for an
acceleration (\ref{con1}) is satisfied, $\xi_{1}$ is automatically negative. In what
follows we shall consider a realistic situation in which the acceleration condition
(\ref{con1}) is imposed. Then the point B is stable when
\begin{eqnarray}
    \label{region}
    -\frac{Q}{Q+\lambda}\leq\Omega_{\vp}<1 \quad\mbox{and}\quad A>0\,.
\end{eqnarray}
The second condition is satisfied if we avoid the ultra-violet instability of quantum
fluctuations \cite{PT}, which is the case for a non-phantom scalar field. From
Eq.~(\ref{scaling4}) the condition $-Q/(Q+\lambda) \leq \Omega_{\vp}$ corresponds to
$-2Q(Q+\lambda) \leq 3(g+g_1)$. This is automatically fulfilled for a non-phantom fixed point
($g+g_1>0$) under the condition (\ref{con1}). The most crucial condition for the stability of
the point B is $\Omega_{\vp}<1$, i.e.,
\begin{eqnarray}
    \label{stacon3}
    g+g_1<\lambda (Q+\lambda)/3\,.
\end{eqnarray}

For a non-phantom fixed point this is not satisfied if $Q<-\lambda$ but can be
satisfied if $Q>\lambda/2$. Hence when $Q>\lambda/2$ there exist stable,
accelerated, and non-phantom fixed points B provided that $\Omega_{\vp}<1$ (whose
condition is actually required to get a viable scaling solution). Note that when
$Q>\lambda/2$ the stability condition given in Eq.~(\ref{stacon2}) has an opposite
equality to that in Eq.~(\ref{stacon3}). Hence the stability of the points A and B
is divided by the border $g+g_1=\lambda (Q+\lambda)/3$, which means that the final
attractor is either the point A or B depending on the values of $Q$ and $\lambda$.
When  a non-phantom scaling solution with positive $Q$ exists in the region
(\ref{region}), it is the only stable attractor point for any $g(Y)$ \footnote{An
exception to this rule exists in the case of the fractional power-law Lagrangian
(\ref{model}) with $0<u<1$, in which a phantom attractor may also coexist. }, so
that scaling solutions have the crucial property of being \emph{global attractors}.

As a last remark, we note that a general form of $g$ could in principle exhibit
several scaling solutions, all of them with the same $x_B$, $g(Y_B)$ and $w_{\rm
eff}$, but with different $y_B$.

\subsection{\label{sub:Kinetic-solutions} Points C and D: Kinetic solutions}

Now we study the second class of solutions of Eq.~(\ref{auto2}), i.e., the case of
$y=0$. These points exist only if $g=g(x^{2}/y^{2})$ is non-singular, i.e., only if
one can expand $g$ in positive powers of $y^{2}/x^{2}$,
\begin{eqnarray}
    g=c_{0}+\sum_{n>0}c_{n}\left(\frac{y^{2}}{x^{2}}\right)^{n}\,,\label{eq:poly2}
\end{eqnarray}
 in which case one has
\begin{eqnarray}
    g_{n}(y\rightarrow0)=0\quad(n>0)\,.\label{eq:poly3}
\end{eqnarray}
 In this case Eq.~(\ref{auto1}) is simply given by
\begin{eqnarray}
\frac{{\textrm{d}}x}{{\textrm{d}}N}=\frac{1}{2}\left(3c_{0}x+\sqrt{6}Q\right)
\left(x^{2}-\frac{1}{c_{0}}\right)=0\,.
\end{eqnarray}
 For $c_{0}=0$ this equation gives no real solutions. For $c_{0}\ne0$
we get the following fixed points:

\begin{itemize}
    \item Point C: a $\varphi$-matter dominated era or $\varphi$MDE (see Ref.~\cite{coupled})
    \begin{eqnarray}
        (x_{C},y_{C})=\left(-\frac{\sqrt{6}Q}{3c_{0}},0\right).\label{MDE}
    \end{eqnarray}
    In this case Eqs.~(\ref{Omedef}) and (\ref{rhowvp}) give
    \begin{eqnarray}
        \;\;\Omega_{\vp}=\frac{2Q^{2}}{3c_{0}}\,,\quad w_{\rm eff}=
        \frac{2Q^{2}}{3c_{0}}\,,\quad w_{\vp}=1\,.\label{phiMDEre}
    \end{eqnarray}
    Hence $\Omega_{\vp}$ and $c_{0}$ always have the same sign. When
    $c_{0}>0$ the solutions is decelerated, and the requirement of the condition
    $\Omega_{\vp}<1$ gives
    \begin{eqnarray}
        |Q|<\sqrt{\frac{3}{2}c_{0}}\,.\label{Qcon}
    \end{eqnarray}
    We note that the $\vp$MDE also corresponds to the same class of
    scaling solutions of point B. In fact setting $g(Y_{C})=c_{0}=-2Q(Q+\lambda)/3$ in
    Eq.~(\ref{scaling1}) and eliminating $\lambda$ in
    Eqs.~(\ref{scaling2})-(\ref{scaling4}), we obtain the results in
    Eq.~(\ref{phiMDEre}). Also, we remark that for $c_{0}>0$, $w_{\rm eff}>0$ and
    therefore dark matter density dilutes as $a^{-3(1+w_{\rm eff})},$ i.e. faster than
    baryons. This ensures that baryons did not dominate in the past. Since during
    acceleration baryons dilute faster than dark energy, we can safely assume that
    baryons never contributed a large portion of cosmic energy. For $c_{0}<0$ this is
    not necessarily true but these cases will be ruled out for other
    reasons presented later.
    \item Point D: pure kinetic solutions
    \begin{eqnarray}
        (x_D,y_D)=(\pm1/\sqrt{c_{0}},0)\,,\label{kine}
    \end{eqnarray}
    which exists only for positive $c_{0}$. In this case we have that matter is absent
    and
    \begin{eqnarray}
        \Omega_{\vp}=1\,,\quad w_{\rm eff}=1\,,\quad w_{\vp}=1\,.
    \end{eqnarray}
\end{itemize}
Let us now consider linear perturbations $\delta x$ and $\delta y$ about the generic
kinetic fixed point $(x,y)=(x_{k},0)$. In this case the $2\times2$ matrix
${\mathcal{M}}$ for perturbations \cite{CLW} is diagonal and its eigenvalues are
given by
\begin{eqnarray}
    \mu_{1} & = & -\frac{3}{2}+\frac{9}{2}c_{0}x_{k}^{2}+\sqrt{6}Qx_{k}\,,\\
    \mu_{2} & = & \frac{3}{2}\left(1+c_{0}x_{k}^{2}-\frac{\sqrt{6}}{3}\lambda
    x_{k}\right)\,.
\end{eqnarray}

Hence in the case of the $\vp$MDE solution the eigenvalues are
$\mu_{1}=(Q^{2}/c_{0})-3/2$ and $\mu_{2}=3/2+Q(Q+\lambda)/c_{0}$. When $c_{0}>0$,
$\mu_{1}$ is negative under the condition (\ref{Qcon}) whereas $\mu_{2}>0$ for the
values of $Q$ satisfying Eq.~(\ref{con1}). This shows that the $\vp$MDE corresponds
to a saddle point for all the relevant cases when $c_{0}>0$. When $c_{0}$ is
negative, it can be a stable point if $Q(Q+\lambda)>3|c_0|/2$.

In the case of pure kinetic solutions (which exist only for $c_{0}>0$) one has
$\mu_{1}=3\pm\sqrt{6/c_{0}}\, Q$ and $\mu_{2}=3\mp\sqrt{6/c_{0}}\,\lambda/2$ for
$x_{k}=\pm1/\sqrt{c_{0}}$. Thus, for $Q>0$, in both cases at least one of the
eigenvalues is positive, which means that the solutions are either unstable nodes or
saddle points depending on the values of $Q$ or $\lambda$. When $Q<0$, the point
$x_{k}=1/\sqrt{c_{0}}$ is stable when $Q<-\sqrt{3c_{0}/2}$ and $\lambda>\sqrt{6\,
c_{0}}$, whereas the point $x_{k}=-1/\sqrt{c_{0}}$ is an unstable node.

\subsection{Summary of fixed points}

\begin{table*}[t]
\begin{center}\begin{tabular}{|c|c|c|c|c|c|c|}
    \hline \textbf{Point} & $\mathbf{x}$ & $\mathbf{y}$ & \textbf{Existence}
    & \textbf{Stability} & $\mathbf{\Omega_{\vp}}$& $\mathbf{w_{\rm eff}}$\tabularnewline
    \hline A &
    $x_{A}$& $(x_{A}^{2}/Y_{A})^{1/2}$& $y_A\neq0 \mbox{ and } \Omega_\vp = 1$&
    \parbox[c][1.0cm]{5cm}{Stable node under \linebreak conditions (\ref{stacon})}& 1 &
    %$-1+\frac{\sqrt{6}\lambda x_{A}}{3}$
    $-1+\sqrt{\frac{2}{3}}\lambda x_A$
    \tabularnewline
    \hline B &
    {\large$\frac{\sqrt{6}}{2(Q+\lambda)}$}&
    $(x_{B}^{2}/Y_{B})^{1/2}$&
    $3(g+g_{1})<(Q+\lambda)\lambda$& \parbox[c][1.0cm]{5cm}{
    Stable node for \linebreak $-\frac{Q}{Q+\lambda}\leq\Omega_{\varphi}<1\;$ and $\;A>0$}&
    {\large$\frac{Q(Q+\lambda)+3(g+g_{1})}{(Q+\lambda)^{2}}$}&
    {\large$-\frac{Q}{Q+\lambda}$}\tabularnewline
    \hline C &
    {\large$-\frac{\sqrt{6}Q}{3c_{0}}$}& $0$& $|Q|<\sqrt{(3c_{0}/2)}$ or $c_0<0$&
    \parbox[c][1.2cm]{5cm}{Saddle point for $\;c_{0}>0$ \linebreak
    Stable node for $\;c_0<0\;$ and $Q(Q+\lambda)>3|c_0|/2$}&{\large$\frac{2Q^{2}}{3c_{0}}$}
    & {\large$\frac{2Q^{2}}{3c_{0}}$}\tabularnewline
    \hline D &
    {\large$\pm\frac{1}{\sqrt{c_{0}}}$}& $0$& $c_{0}>0$&
    \parbox[c][.9cm]{5cm}{Unstable node or saddle for $Q>0$ \linebreak
    Stable node for $Q<-\sqrt{3c_0/2}$}& $1$& $1$ \tabularnewline
    \hline
\end{tabular}\end{center}

\caption{The properties of critical points for the Lagrangian density (\ref{geLag2})
in the presence of a pressureless dust ($w_{m}=0$). Specifying the form of $g(Y)$,
$x_{A}$ and $Y_{A}$ are determined by solving Eq.~(\ref{gre1}) whereas $Y_{B}$ is
known by Eq.~(\ref{scaling1}). The kinetic fixed points C and D exist when $g$ is
given by Eq.~(\ref{eq:poly2}). \label{crit}}
\end{table*}

In Table\,\ref{crit} we summarize the property of fixed points for the Lagrangian
density (\ref{geLag}). The scalar-field dominated fixed point A and the scaling
solution B exist for any form of $g(Y)$ as long as they satisfy the condition of
existence given in the Table\,\ref{crit}. Both fixed points can be used for
late-time acceleration, since the effective equation of state $w_{\rm eff}$ can be
smaller than $-1/3$ depending upon the values of $Q$ and $\lambda$.
For a non-phantom case the final attractor is either A or B depending
on the values of $Q$ and $\lambda$.
The scaling solution B is a global attractor provided that the
condition (\ref{region}) is satisfied.

The existence of the kinetic fixed points C and D depends on the form of the scalar-field
Lagrangian. They appear when $g$ is expanded into positive powers of $y^{2}/x^{2}$, i.e.,
Eq.~(\ref{eq:poly2}). An ordinary scalar field with an exponential potential ($g=1-c/Y$)
belongs to this class, while for instance a dilatonic ghost condensate model \cite{PT}
($g=-1+c\,Y$) does not. The fixed point C corresponds to a saddle point for $c_{0}>0$ with
$\Omega_{\vp}=w_{\rm eff}=2Q^{2}/3c_{0}$. Hence one can have a temporary scalar-field matter
dominated era ($\vp$MDE) in the presence of the coupling $Q$. We note that $\Omega_{\vp}<0$
when $c_{0}$ is negative. The fixed point D appears only for positive $c_{0}$ and corresponds
to $\Omega_{\vp}=1$ with no acceleration ($w_{\rm eff}=1$). Hence this is neither viable for
the matter-dominated era nor for the dark energy dominated era and it will not be considered
further.

\section{Can we have two scaling regimes?}

\label{twoscaling}

As we anticipated in the Introduction, we search now for the occurrence of a two-stage
cosmology: a decelerated matter epoch and an accelerated scaling regime. This amounts to
searching for two distinct fixed points for the same set of parameters $\{Q,\,\lambda\}$.
Clearly the matter point has to be a saddle point in order to give way to the final
accelerated stable attractor. Since in general during the matter epoch there will be a
non-negligible contribution of the scalar field, this point is, in general, a scaling point.
Therefore we search for two subsequent scaling regimes. It is in principle possible to obtain
an approximate matter epoch without an associated fixed point but this would require a fine
tuning of the initial condition, so we exclude this possibility here.

As we have shown in the previous section there are two possibilities which lead to
an accelerated expansion at late times--using either the scalar-field dominated
fixed point A or the scaling solution B. The $\vp$MDE fixed point C appears prior to
the accelerated epoch for the models given by Eq.~(\ref{eq:poly2}). For an ordinary
scalar field with an exponential potential it was found that the $\vp$MDE is
followed by the attractor point A \cite{coupled} (or by point B but in this case
without acceleration). In this case the present universe ($\Omega_{\vp}\simeq0.7$)
would be finally dominated by the energy density of the scalar field
($\Omega_{\vp}=1$). Conversely, if the present accelerated universe corresponds to a
scaling attractor B, it was shown that the matter dominated epoch, if any, is not
sufficiently long to form large-scale structure. This is associated with the fact
that we require a large coupling $Q$ to obtain an accelerated scaling attractor, but
in this case the solution quickly approaches the attractor after the end of a
radiation era since there is no saddle matter point. Hence one can not have two
scaling regimes (the decelerated point C and the accelerated point B) at the same
time for the standard scalar field with an exponential potential.

In this section we investigate whether two scaling solutions can be realized for the
general Lagrangian (\ref{geLag}) with $g$ given by Eq.~(\ref{eq:poly}). We scan all
the parameter space \{$Q\,,\lambda$\} in search of a successful scaling cosmology.
Since the cosmological dynamics is different depending on the sign of $c_{0}$, we
shall consider three cases (i)~$c_{0}>0$, (ii)~$c_{0}<0$ and (iii)~$c_{0}=0$
separately. We shall also look into the alternative, fractional power law Lagrangian
given by
\begin{eqnarray}
    g(Y)=c_{0}-c\, Y^{-u}\,,\label{model}
\end{eqnarray}
 where $u$, as opposed to $n$, is not limited to integer values.

\subsection{Case of $c_{0}>0$}

The function $g$ given in Eq.~(\ref{eq:poly}) is composed by positive and negative
powers of $Y$. We shall first show that the case of positive powers of $Y$ in
Eq.~(\ref{eq:poly}) is not cosmologically viable and then proceed to the case of
negative values of $n$.

\subsubsection{Positive powers of $Y$}

Let us first consider the function $g$ given by
\begin{eqnarray}
    g=c_{0}+\sum_{n<0}c_{n}Y^{-n}\,.\label{positive}
\end{eqnarray}
In this case all the critical points with $y=0$ disappear because of the
singularity. Then, the only possibilities giving rise to a matter-dominated phase
corresponds to either $x=0\,$ or $\,g=g_1=0$ (see Eqs. (\ref{Omedef}) and
(\ref{weffex})), for which indeed $w_{\rm eff}=0$ and $\Omega_{\varphi}=0$. For
$x\rightarrow 0$, however, one has $g\rightarrow c_{0}$ and $g_{1},g_{2}\rightarrow
0$ from Eq.~(\ref{positive}). Then it is immediate to see that
${\textrm{d}}x/{\textrm{d}}N \, (x\rightarrow0) = -\sqrt{6}Q/(2c_{0})$ from
Eq.~(\ref{auto1}), so we do not have a fixed point unless of course $Q=0$ (in which
case the scaling solution B is not accelerated, as can be seen from
Eq.~(\ref{con1})). If $g=g_1=0$, the  situation is the same and again we only have
fixed points when $Q=0$. Thus we do not have a successful cosmological scenario for
the function given by Eq.~(\ref{positive}).
Note also that for the model $g=c_{0}-cY^{-u}$ with negative $u$
the scalar-field energy fraction $\Omega_{\varphi}$
for the point B has to be negative from Eq.~(\ref{eq:ophiB}) when the condition
(\ref{con1}) for an acceleration is imposed.

\subsubsection{Negative powers of $Y$}

Since we have seen that all positive powers of $Y$ in the general polynomial form of
$g$ are discarded, let us then focus on polynomials with negative power of $Y$,
i.e.,
\begin{eqnarray}
    g=c_{0}+\sum_{n>0}c_{n}Y^{-n}\,.\label{negative}
\end{eqnarray}
 We will prove that when $\,c_n\neq0\,$ for $\,n\geq1\,$ it is impossible to have two viable
scaling regimes which satisfy observational constraints.

From Eqs.~(\ref{MDE}) and (\ref{scalingB}) we see that the $\vp$MDE decelerated solution C and
the accelerated point B have always opposite signs of $x$. In fact, requiring the acceleration
at B, we have either $Q>\lambda/2>0$ or $Q<-\lambda<0$. In the former case one has $x_{B}>0$
and $x_{C}<0$, whereas in the latter case $x_{B}<0$ and $x_{C}>0$. However, the function $g$
given in Eq.~(\ref{negative}) is singular at $x=0$ (except for power-laws $u\leq1$, see
below), which implies that the sequence of the solution from C to B is prevented. In what
follows we will provide a more detailed analysis for the possibility of getting two scaling
regimes.

Let us first consider a single power-law function of $g(Y)$ given in
Eq.~(\ref{model}). In the limit $x\rightarrow0$ with a nonzero value of $y$ (which
can be $y\ll1$), the $gx^{2}y$ term on the r.h.s. of Eq.~(\ref{auto2}) exhibits a
divergence for $u>1$ together with a divergence of the $gx^{3}$ term on the r.h.s.
of Eq.~(\ref{auto1}) for $u>3/2$. In fact, for $u\neq1$ we have that
\begin{equation}\label{eq:xsingularity}
    \bigg|\frac{\textrm{d}y/\textrm{d}N}{\textrm{d}x/\textrm{d}N}\bigg|_{x\rightarrow0}
    \rightarrow \infty \,.
\end{equation}
Hence when $u\neq1$ the solutions cannot pass the line given by $x=0$. Since the
signs of $x_{B}$ and $x_{C}$ are always different, it is inevitable to hit this
singularity for $u>1$ if the solutions move from the $\vp$MDE point C to the scaling
solution B. This shows that a sequence of solutions from C to B is forbidden because
of the singularity at $x=0$.

\begin{figure}
\includegraphics[width=7cm]{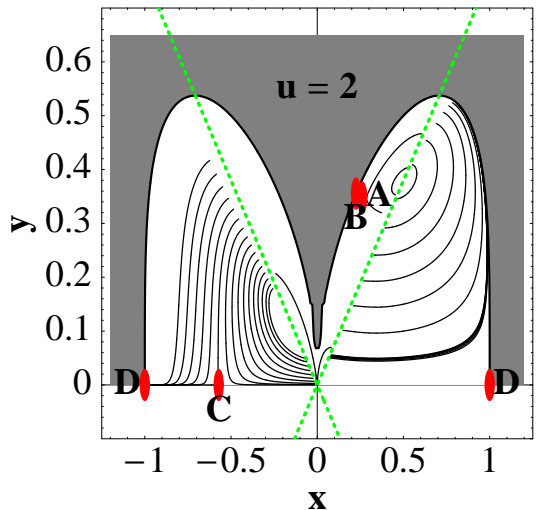}
\caption{Phase space for the model (\ref{model}) with $u=2$, $c=c_0=1$, $\lambda=4$ and
$Q=0.7$ together with the fixed points A, B, C and D. Here and in the following figures the
gray area represents the region where $\Omega_{\varphi}>1$. The dotted line corresponds to the
singularity given by (\ref{ysingularity}) at which the speed of sound diverges.} \label{n2}
\end{figure}

In Fig.~\ref{n2} we plot a phase space for the model~(\ref{model}) with $u=2$, $c=c_0=1$,
$\lambda=4$ and $Q=0.7$ together with the fixed points of the system. The phase space is
characterized by another singularity in Eq.~(\ref{auto1}), associated with the divergence of
the speed of sound. This appears when the quantity, $A^{-1}=c_{0}-c(u-1)(2u-1)Y^{-u}$, becomes
equal to zero, i.e.,
\begin{eqnarray}
    y=\pm\left(\frac{c_{0}}{c(u-1)(2u-1)}\right)^{1/2u}x\,.\label{ysingularity}
\end{eqnarray}
For positive $c$, it exists for $\,u>1\,$ or $\,0<u\leq1/2\,$ but disappears for
$\,1/2<u\leq1\,$. When $\,c<0\,$ the converse is true.

\begin{figure}
\includegraphics[width=7cm]{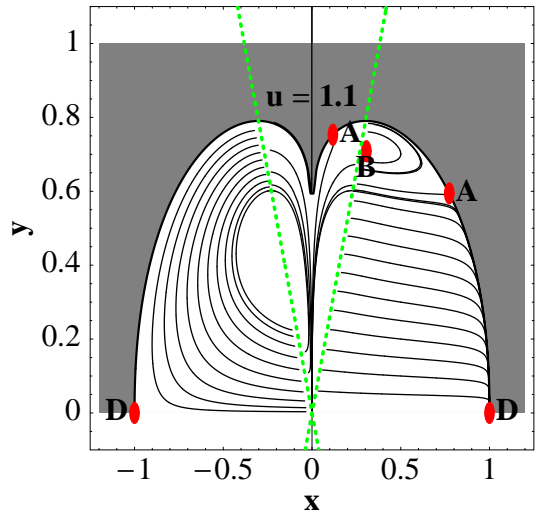}
\caption{Phase space for the model (\ref{model}) with $u=1.1$, $c=c_0=1$, $\lambda=2$ and
$Q=2$ together with the fixed points A, B and D (point C lies in the $\Omega_{\varphi}>1$
region). The dotted line corresponds to the singularity given by (\ref{ysingularity}) at which
the speed of sound diverges.} \label{n11}
\end{figure}

We note that for the model (\ref{model}) the fixed point B corresponds to
\begin{equation}
    x_{B}=\frac{\sqrt{6}}{2(Q+\lambda)}, \;\;\;
    y_{B}=\left(\frac{2Q(Q+\lambda)+3c_{0}}{3c}\right)^{1/2u}\!\! x_{B}\,.
\end{equation}
When the condition (\ref{con1}) for acceleration is satisfied, we require $c>0$ for the
existence of the point B. Then in what follows, we shall only consider the case of positive
$c$. When $u>3/2$ the point B does not satisfy the condition $A>0$, i.e.,
$y<\left(\frac{c_{0}}{c(u-1)(2u-1)}\right)^{1/2u}|x|$. This can be checked in Fig.~\ref{n2} in
which the scaling solution B exists in the region $A<0$. When $1<u<3/2$, it is possible to
obtain positive values of $A$. However we still need another condition:
$\Omega_{\vp}^{(B)}<1$, which gives a more severe constraint. Unless $u$ is close to 1, it is
not easy for the critical point B to fulfill the conditions $A>0$ and \linebreak[4]
$\Omega_{\vp}^{(B)}<1$. One example satisfying these conditions is $u=1.1$, $Q=2$, $\lambda=2$
with $c_{0}=c_{1}=1$, as plotted in Fig.~\ref{n11}. In this case, however, the $\vp$MDE point
exists in the region $\Omega_{\vp}^{(C)}>1$. More importantly the trajectories can not move
from the point C to B because of the singularities at $x=0$ and also at $A^{-1}=0$.

The above discussion shows that when $u>1$ one can not realize two scaling regimes.
On the other hand, the $u=1$ case (an ordinary scalar field with an exponential
potential) is free from both singularities at $x=0$ and $A^{-1}=0$. This case
however has been already ruled out as a successful cosmological model
\cite{coupled}. The argument is as follows; we leave $u$ as a free parameter to
clarify some interesting aspects of the more general case. The relevant quantities
for the fixed points B and C are given in Eqs.~(\ref{scaling2}), (\ref{phiMDEre})
and by the following relation:
\begin{eqnarray}
     &  & \Omega_{\vp}^{({\rm B})}=\frac{(2u-1)Q(Q+\lambda)+3uc_{0}}{(Q+\lambda)^{2}}\,.
    \label{eq:ophiB}
\end{eqnarray}
Let us impose the observational constraints that during the phase C: $\Omega_{\vp}^{({\rm
C})}<1$ and during the phase B: $w_{\rm eff}^{({\rm B})}<-1/3$ and $\Omega_{\vp}^{({\rm
B})}<1$ (notice that the supernovae observed value $w_{\rm DE}$ is in reality defined through
the standard Friedmann equation so it cannot be directly used here; we have shown elsewhere
\cite{agp} that the best value for $w_{\rm eff}^{({\rm B})}$ is in fact around $-0.6\pm0.1$).
In reality observations require quite more stringent constraints than this. For instance,
supernovae observations constrain $\Omega_{\vp}^{({\rm B})}=0.7\pm0.2$ and $w_{\rm eff}^{({\rm
B})}<-0.6\pm0.1$. Moreover, too much amount of dark energy during the phase C leads to a weak
growth of perturbations and serious conflicts with the CMB, so a conservative limit would be
$\Omega_{\vp}^{({\rm C})}<0.2$ (see e.g., Ref.~\cite{coupled}).

\begin{figure}
\includegraphics[width=7cm]{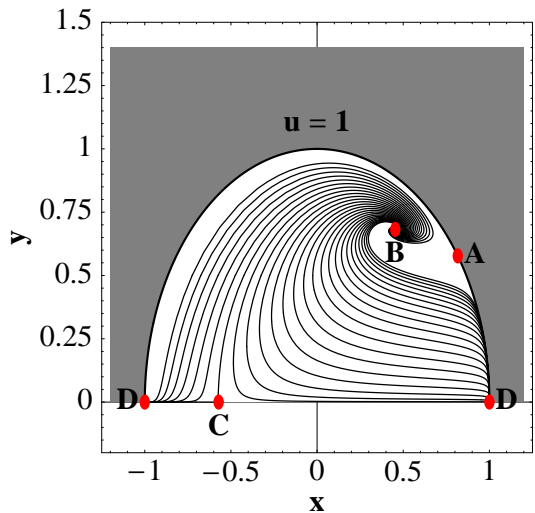}
\caption{Phase space for the model (\ref{model}) with $u=1$, $c=c_0=1$, $\lambda=1.54$ and
$Q=1.02$ together with the fixed points A, B, C and D.}\label{n1}
\end{figure}

The condition for the acceleration of the scaling solution B ($w_{\rm eff}^{({\rm B})}<-1/3$)
requires either the condition $\,Q>\lambda/2>0\,$ or $\,Q<-\lambda<0$. In the latter case it
is easy to find that $\Omega_{\vp}^{(\rm{B})}$ becomes larger than 1 for $u>1$. In the former
case the condition $3Q>Q+\lambda$ gives
\begin{equation}
    \Omega_{\vp}^{({\rm B})}>\frac{uc_{0}}{3Q^{2}}+\frac{2u-1}{3}\,,
\end{equation}
 while the condition $|Q|<\sqrt{3c_{0}/2}$ for the existence of the
$\varphi$MDE implies
\begin{equation}
    \Omega_{\varphi}^{({\rm B})}>\frac{8u-3}{9}\,.
\end{equation}
This shows explicitly that for any $u\geq3/2$ the existence/acceleration of the point B is in
contradiction with the existence of the point C. When $u=1$ it is possible to have two scaling
solutions C and B, but we do not get values smaller than $w_{\rm eff}^{({\rm B})}=-0.4$,
$\Omega_{\vp}^{({\rm B})}=0.9$ and $\Omega_{\vp}^{({\rm C})}=0.7$. This certainly excludes the
$u=1$ case from the range of viable cosmological models. The values of the parameters that
match this limit are $\lambda=1.54$ and $Q=1.02$. The phase space plot in this case is
presented in Fig.~\ref{n1}, from which it is clear that the $\vp$MDE fixed point C is indeed
followed by the scaling solution B without singularities, although such solutions are not
cosmologically viable. When $1<u<3/2$ one can not satisfy the observational constraints
either, in addition to the impossibility of reaching the point B from C. In Fig.~\ref{n11} we
plot a phase space for the model (\ref{model}) with $u=1.1$, $\lambda=2$ and $Q=2$ together
with the fixed points of the system.

We have thus shown that it is not possible to obtain two ideal scaling regimes for
the function (\ref{model}) with $u\geq1$. Now we extend this proof to a polynomial
form of $g(Y)$ with negative powers. The problem as we have seen is mainly
associated with the fact that two scaling solutions B and C are separated by
singularities at $x=0$ and \linebreak[3] $A^{-1}=0$. The latter can disappear by
considering the sum of the powers given in Eq.~(\ref{negative}) with the
adjustment of the coefficients $c_{n}$. However if the polynomial includes any power
$n$ larger than 1, this leads to a singularity at $x=0$ even when the singularity at
$A^{-1}=0$ is not present. Hence if the function $g$ possesses at least one term
whose power $n$ is larger than 1, the polynomials (\ref{negative}) are excluded as
an ideal scaling cosmology. This completes our proof of the impossibility of
obtaining two scaling solutions in the case of negative powers of $Y$ and positive
$c_{0}$.

Let us conclude this subsection with a brief discussion of the case $u<1$ in
Eq.~(\ref{model}). When $u<1$ the line $x=0$ is no longer singular; however, Eq.
(\ref{eq:xsingularity}) still holds and the phase space is again separated into
positive and negative abscissa subspaces. Moreover, the singularity at $A^{-1}=0$
disappears for $1/2<u<1$ (remember we are only interested in $c>0$).
%We can thus have a matter-dominated era in the region $x<0$ followed by the scaling
%solution B with $x_B>0$ and vice-versa.
% I think negative x case is followed by a phantom point A.
In this case it is also possible to have another nearly matter-dominated phase. This
corresponds to a situation in which one takes a limit \linebreak[4] $x\rightarrow0$ with a
nonzero but small $y$ in Eqs.~(\ref{auto1}) and (\ref{auto2}). Then we can have a
matter-dominated era in the region $x>0$ followed by the scaling solution B with $x_{B}>0$
(when $Q$ is positive). This situation is similar also in the case $0<u\leq1/2$; as long as
the system is in the region $x>0$ and $A>0$ initially, the solutions reach the scaling
attractor B without any singularity. It is interesting to remark that when $u<1$ and $c>0$,
every class A points with $x_{A1}>0$ is accompanied by a second point A: a phantom attractor
($x_{A2}<0$). Hence, it can happen that the $\vp$MDE be followed by a point A without a
singularity even for $Q>0$ ($x_C<0$). The $u<1$ ``fractional Lagrangians'' are then promising
but clearly for these models to work there are several other observational and theoretical
issues that should be considered and we leave them to a future work.

\subsection{Case of $c_{0}<0$}

We shall next consider the case of negative $c_{0}$. The positive powers of $Y$
given in Eq.~(\ref{positive}) are excluded as viable cosmological scenarios by
arguments similar to those presented in the previous subsection. Then let us focus
on the negative powers of $Y$ given in Eq.~(\ref{negative}). In this case one has
the $\vp$MDE solution (\ref{MDE}) with a negative $\Omega_{\vp}$ {[}see
Eq.~(\ref{phiMDEre}){]}.

In addition to the fact that this may be unphysical, we are also faced with another
problem to obtain two scaling regimes C and B. Since the $\vp$MDE satisfies the
condition (\ref{eq:poly3}) one has $A^{-1}=g+5g_{1}+2g_{2}=c_{0}$, which means that
$A$ is negative. On the other hand in order to get a stable scaling solution~B with
$0\leq\Omega_{\vp}<1$, we require $A$ positive. Then, to reach the point B from the
point C, one needs to cross either the singularity at $A^{-1}=0$ (which is not
allowed) or go through $A=0$ ($x=0$). The latter can only be accomplished in the
alternative model~(\ref{model}) with $u=1$. Therefore one can not realize a good
scaling cosmology when $c_{0}$ is negative.

\subsection{Case of $c_{0}=0$}

The case $c_{0}=0$ is also easy to dispose of. In fact in this case there are no kinetic
solutions and therefore no matter eras with $y=0$ {[}see Eqs.~(\ref{MDE}) and (\ref{kine}){]}.
One can have a matter era also for $x\rightarrow\infty$ with $n\geq1$ or for $x\rightarrow0$
with fractional powers less than one. However in both cases $A$ is singular and therefore there
are no fixed points. Finally, if both $y$ and $x$ go to zero so that $Y=\mbox{const.}$, then
one can verify that ${\textrm{d}}x/{\textrm{d}}N$ does not vanish and therefore the point
$(0,0)$ is not a solution.

This completes our proof. Although the discussion has been rather long and technical, the conclusion
is straightforward: we have  shown that no cosmologically viable scaling solutions exist for the general class of integer polynomial field Lagrangians with variable coupling.

\section{Conclusions}

In this paper we have addressed a number of interesting aspects of cosmological
scaling solutions and derived the following results.

\begin{itemize}
\item We have identified the most general form of second-order scalar field
Lagrangian given in Eq.~(\ref{eqn:solution-for-p}) with a coupling to matter that is a
completely arbitrary function of $\vp$ (but does not depend on $X$) under
the condition that the system exhibits scaling solutions. This is the generalization of the
works \cite{PT,TS} in which a similar form of Lagrangian was obtained in the case of a
constant coupling.
\item We have classified the phase space topology for the scaling Lagrangian
and obtained four classes of fixed points: (A) scalar-field dominated points with
$\Omega_{\vp}=1$, (B) scaling solution with \linebreak[3]
$\Omega_{\vp}=-Q/[w_{\vp}(Q+\lambda)]$, (C)~a $\vp$MDE solution, and (D)~pure kinetic
solutions. Points of the first two classes may exist for any scaling Lagrangian and can lead
to an accelerated expansion. The accelerated scaling attractor B, when it exists in the region
$-Q/(Q+\lambda) \leq\Omega_{\vp}<1$,
is the only global attractor apart from the case in which another
phantom attractor A is present. The points C and D appear when the function $g$ can be
expanded in the form (\ref{eq:poly2}). The $\vp$MDE solution C is another scaling solution
(always decelerated in the cases of interest).
\item We have addressed the possibility of finding a sequence of matter
and scaling acceleration and found that this is {\it impossible} for any scaling
Lagrangian which can be approximated as a polynomial with both positive and negative
integer powers of its argument $Y$. This is essentially due to the fact that a
scaling Lagrangian is always singular either along the $x$-axis or the $y$-axis of
the phase space, thereby either preventing the matter-dominated era or isolating the
region with a viable matter era from the region where the scaling acceleration
occurs.
\end{itemize}
It is rather remarkable that the sequence of two scaling regimes cannot be realized for such a
vast class of scalar-field Lagrangians (although, to be fair, we did not investigate
thoroughly the consequences of Eq.~(\ref{eq:poly}) having infinite terms). This underlines how
difficult it is to solve the problem of coincidence: although cosmological scaling solutions have
been studied for over a decade now, no successful case has been identified and this paper
shows that even a large generalization of the models does not help. The search for a good
scaling cosmology is not over yet, though. In fact we have also shown that a possible
exception exists in the $0<u<1$ sector of the function $g(Y)$ given in Eq.~(\ref{model}). A
detailed investigation of this type of fractional Lagrangian is underway.

\section*{ACKNOWLEDGEMENTS}

L.~A. thanks Gunma National College of Technology for hospitality and JSPS for support. M.~Q.
and I.~W. want to thank INAF/Osservatorio Astronomico di Roma for hospitality. M.~Q. is
completely and I.~W. is partially supported by the Brazilian research agency CNPq. S.~T. is
supported by JSPS (Grant No.\,30318802).

\section*{Appendix A - Derivation details}

In order to solve Eq. (\ref{eqn:gen-master-eqn}), we first rewrite it using as a field
$\lambda$ times Eq. (\ref{eqn:def-of-psi}). That is, $\partial/\partial\vp=\lambda\,
Q(\vp)\,\partial/\partial\psi$:
\begin{equation}
    \frac{\partial\ln p}{\partial\ln
    X}\left[1+\frac{2}{Q(\vp)}\frac{{\textrm{d}}Q(\vp(\psi))}{{\textrm{d}}\psi}\right]-
    \frac{\partial\ln p}{\partial\psi}=1.\label{eqn:master_eq_rewritten}
\end{equation}
We then decompose $p(X,\vp)$ into
\begin{equation}
    \ln p(X,\vp)=-\psi+\ln f(X,\psi),\label{eqn:solving:decomp}
\end{equation}
thus arriving at
\begin{equation}
    \frac{\partial\ln f}{\partial\ln
    X}\left[1+\frac{2}{Q(\vp(\psi))}\frac{{\textrm{d}}Q(\vp(\psi))}{{\textrm{d}}\psi}\right]-
    \frac{\partial\ln f}{\partial\psi}=0.\label{eqn:master-eq-rewritten-2}
\end{equation}
This last equation can be solved by Fourier analysis:
\begin{equation}
    \ln f(X,\psi)\equiv\frac{1}{\sqrt{2\pi}}\int e^{i\omega\ln
    X}F(\omega,\psi)\,{\textrm{d}}\omega.\label{eqn:solving:fourier}
\end{equation}
Equation (\ref{eqn:master-eq-rewritten-2}) then becomes (with $Q_{\psi}\equiv Q\circ\vp$,
where we use $\circ$ to denote a composite function)
\begin{equation}
    i\omega F
    \left[1+\frac{2}{Q_{\psi}}\frac{{\textrm{d}}Q_{\psi}}{{\textrm{d}}\psi}\right]=
    \frac{\partial F}{\partial\psi},\label{eqn:solving:fourier-2}
\end{equation}
which has as solution
\begin{equation}
    \ln F=\int^{\psi}i\omega\; \left[1+\frac{2}{Q_{\psi}(z)}
    \frac{dQ_{\psi}(z)}{dz}\right]dz +B(\omega), \label{eqn:solving:fourier-3}
\end{equation}
where $B(\omega)$ is an arbitrary function. Undoing the Fourier transformation, we get
\begin{eqnarray}
    \ln f & = & \frac{1}{\sqrt{2\pi}}\int d\omega\; B(\omega)\exp\Bigg\{ i\omega\Bigg[
    \ln X\;+\;\;\;\nonumber \\
    &  & +\underbrace{\int^{\psi}\hspace{-.1cm}\left(1+\frac{2}{Q_{\psi}(z)}
    \frac{dQ_{\psi}(z)}{dz}\right)dz}\Bigg]\Bigg\}.\;\label{eqn:solving:fourier-4}\\
    &  & \hspace{.4cm}=\psi+2\ln Q_{\psi}(\psi)+\mbox{const.}\nonumber
\end{eqnarray}

Finally, recalling (\ref{eqn:solving:decomp}) and noting that the composite function
$Q_{\psi}\circ\psi=Q$, we arrive at
\begin{equation}
    p(X,\vp)=e^{-\psi(\vp)}\;\tilde{g}\left(X\; e^{\psi(\vp)}\;
    Q^{2}(\vp)\right),\label{eqn:solving:last}
\end{equation}
from which Eq.~(\ref{eqn:solution-for-p}) follows immediately by a redefinition of the
arbitrary function $\tilde{g}$ and by setting $\psi_{\rm new} \rightarrow \psi_{\rm
old}/\lambda$. Notice that the arbitrariness of the function $B$ is absorbed into that of
$\tilde{g}$.

\section*{Appendix B - More general cosmological background}

For completeness we also derive the scaling Lagrangian in an effective FRW equation
which is given by
\begin{equation}
    H^{2}=\beta_{q}^{2}\rho_{\rm T}^{q}\,,\label{Hubble2}
\end{equation}
 where $\beta_{q}$ and $q$ are constants. General Relativity, Randall-Sundrum
braneworlds \cite{brane}, Gauss-Bonnet braneworlds \cite{GB} and Cardassian
Cosmology \cite{Carda} correspond to $q=1$, $q=2$, $q=2/3$ and $q=1/3$,
respectively. The equations (\ref{eqn:conserv-eqn-field}) and
(\ref{eqn:conserv-eqn-m}) are unchanged even for the background (\ref{Hubble2}). The
definition of $\Omega_{\vp}$ and $\Omega_{m}$ are modified as
\begin{equation}
    \Omega_{\vp}\equiv\frac{\rho_{\vp}}{(H/\beta_{q})^{2/q}}\,,\quad\Omega_{m}\equiv
    \frac{\rho_{m}}{(H/\beta_{q})^{2/q}}\,,
\end{equation}
 which satisfies $\Omega_{\vp}+\Omega_{m}=1$ from Eq.~(\ref{Hubble2}).

While Eq.~(\ref{eqn:drho-dN}) holds for $q\neq1$ as well,
Eq.~(\ref{eqn:X-proportional-to}) is subject to the change:
\begin{equation}
    2\,X=H^{2}\left(\frac{{\textrm{d}}\vp}{{\textrm{d}}N}\right)^{2}\;\;\propto\;\;
    \frac{\rho_{\vp}^{q}}{Q^{2}}\;\;\propto\;\;\frac{p^{q}(X,\vp)}{Q^{2}}.
\end{equation}
 Then we obtain the following master equation
\begin{equation}
    q\left[1+\frac{2}{q\lambda\,
    Q^{2}}\frac{{\textrm{d}}Q(\vp)}{{\textrm{d}}\vp}\right]\frac{\partial\ln
    p}{\partial\ln X}-\frac{1}{\lambda Q}\frac{\partial\ln
    p}{\partial\vp}=1,\label{eqn:gen-master-eqn2}
\end{equation}
 where $\lambda$ is defined in Eq.~(\ref{eqn:def-lambda}). The
integration of this equation gives
\begin{equation}
    p(X_{\psi},\psi)=\big(XQ^{2}(\vp)\big)^{1/q}\; g\left(XQ^{2}(\vp)\,
    e^{q\lambda\psi}\right),\label{eqn:2nd-solution-for-p}
\end{equation}
 where $\psi$ is defined in Eq.~(\ref{eqn:def-of-psi}). For constant
$Q$ this reproduces the result obtained in Ref.~\cite{TS}.

\section*{Appendix C - Fixed points for $z\neq0$}

We shall derive here the fixed points for $z\neq0$. In this case one has $z^{2}=1-3gx^{2}$
from Eq.~(\ref{auto3}). Then from Eq.~(\ref{weffex}) the effective equation of state always
corresponds to $w_{\rm eff}=1/3$, which means that the scale factor evolves as $a\propto
t^{1/2}$. Hence we can not use the fixed points with $z\neq0$ to get a matter dominated era or
an accelerated expansion. By substituting the relation $z^{2}=1-3gx^{2}$ for
Eq.~(\ref{auto2}), we find the following two cases: (i) $y=0\:$ and $\,$(ii)
$x=4/(\sqrt{6}\lambda$).

The case (i) is similar to the kinetic solutions discussed in Sec.~IV. Then by considering the
function $g(Y)$ given in Eq.~(\ref{eq:poly2}) one gets
${\textrm{d}}x/{\textrm{d}}N=-x(1+\sqrt{6}Qx)=0$, which gives $x=0$ or $x=-1/\sqrt{6}Q$. Thus,
for $y=0$ we have two fixed points: (a) $(x,y,z)=(0,0,1)$ and (b)
$(x,y,z)=(-1/\sqrt{6}Q,\,0\,,\sqrt{1-c_{0}/2Q^{2}})$. The point (a) corresponds to a standard
radiation dominated era with $\Omega_{\vp}=0$, whereas for the point (b) there is an energy
fraction of the scalar field given by $\Omega_{\vp}=c_{0}/6Q^{2}$.

In the case (ii) we have $x=4/\sqrt{6}\lambda$ and $z^{2}=1-8g/\lambda^{2}$, while $y$ is only
determined by the specific form of $g$ and could as well be zero. From Eq.~(\ref{auto1}) we
obtain the relation $A(\lambda+4Q)(g-g_{1})=0$, which leads to three different fixed points:
(c) $g=g_1$, (d) $\lambda=-4Q$, and (e) $A=0$. Solving $g=g_{1}$ by
specifying the form of $g$, we obtain the value $y=y_c(\neq0)$, i.e., the fixed point (c)~$(x,y,z)=(4/\sqrt{6}\lambda,
\,y_c\,, \sqrt{1-8g/\lambda^{2}})$. The special case where $\lambda=-4Q$ (i.e., $Q<0$) is an
interesting one because $y_d$ and $z_d$ are not specified even after the form of $g$ is given.
In fact, given the form of $g$, the \emph{critical curve} (d) is found by solving the relation
$z_d = \sqrt{1-8\,g\big(8/(3 \lambda^2 y_d^2)\big)}$. Finally, the (e) point demands
that $y$ is zero, otherwise since $x$ is finite, $g$ would be finite and
$A=(g+5g_{1}+2g_{2})^{-1}$ would not vanish. But then $z$ is either imaginary or infinite, so
this fixed point is never realized.

\end{document}